\documentstyle[preprint,aps,epsfig]{revtex}
\tighten
\begin{document}

\draft

\title{New parameterization of the trinucleon wavefunction and its
application to the $\pi\, ^3$He scattering length}

\author{V. Baru$^{1,2}$, J. Haidenbauer$^1$, C. Hanhart$^1$, and J. A.
Niskanen$^3$}

\address{$^1$Institut f\"ur Kernphysik, Forschungszentrum
J\"ulich, D-52425 J\"ulich, Germany, \\
$^2$Institute of Theoretical and Experimental Physics,
117259, B. Cheremushkinskaya 25,\\ Moscow, Russia \\
$^3$Department of Physical Sciences, PO Box 64,
FIN-00014 University of Helsinki, Finland}




\maketitle

\begin{abstract}
We present a new parameterization of the trinucleon wavefunction.
As a novel feature a separable parameterization for the complete wavefunction
is given. In this way
any calculation that considers two-body currents only is largely
simplified. To demonstrate this we calculate the $\pi\, ^3$He
scattering length in chiral perturbation theory.
We find reasonable agreement with experimental values
inferred from data on level shifts in pionic $^3$He bound states.
The relevance of the
$\pi$-triton system for an alternative determination of
the $\pi N$ scattering lengths is discussed.
\end{abstract}
\vskip 1cm \noindent

\pacs{PACS: 25.80.Ls, 25.80.Hp, 25.10.+s, 21.45.+v}

\vskip 2cm


\newpage

\section{Introduction}

In spite of the importance of three- and four-nucleon systems
as a bridge between the deuteron and heavier nuclei, there is an
appalling shortage of serious theoretical work done on meson-nuclear
physics with these.
One stumbling stone for calculations of meson-nuclear
interactions by "intermediate energy physicists" is the need for
at least the trinucleon ($^3$He or $^3$H) Faddeev wave functions,
which normally exist in the form of numerical tables computed by a different
society, low energy physicists. The necessity of a more accessible
form for nonspecialists
was realized a long time ago by Hajduk {\it et al.} \cite{hgs},
who presented the Faddeev amplitudes of the trinucleon wave function
for different channels and
particle permutations as separable analytical forms. The aim
of this work is to improve the parameterization of Ref. \cite{hgs}
in three respects:
\begin{itemize}
\item The parameterization is given for trinucleon wave functions derived
from two modern nucleon--nucleon interactions, namely the
CD Bonn \cite{bonn} and the Paris potential \cite{paris}.
\item We give a separable expansion for the full antisymmetrized wave function.
Previous
works \cite{hgs} parameterized the Faddeev components only.
\item The trinucleon wave function is not separable in all partial waves over
the full momentum range. Therefore we include an additional term in the
parameterization that allows the inclusion of correlations. Especially for
the pair wave functions where the $NN$ pair is in a
$\, ^3d_1$ state this turns out to be crucial.
\end{itemize}

In the early 1990's Weinberg argued, that as long as we restrict
ourselves to interactions with pions only, three and more body
interactions are suppressed. Thus, for those calculations,
carried out either in chiral perturbation theory or within
a phenomenological
approach, all that is needed is an appropriate parameterization of
the nuclear wave function, which contains the momentum distribution
of one active pair only. All remaining degrees of freedom can be
integrated/summed separately. If this parameterization is given
for the full antisymmetrized wavefunction, to do the actual
calculation of a nuclear matrix element with a two-body operator
will be just as complicated as the evaluation of the matrix element
on the deuteron. In this paper we demonstrate that this program
can be easily carried out for the example of the $\pi\, ^3$He scattering length.

The paper is organized in the following way:
Some basic information about the trinucleon wave function and
details of the procedure how the wave function is parameterized by means
of simply analytical functions are given in the next section. In the
third section the usefulness of the new parameterization is demonstrated
by an explicit calculation of the $\pi \, ^3$He scattering length within
chiral perturbation theory. The paper ends with a short summary.

\section{Parameterization of the trinucleon wave function}
The full antisymmetric wave function of a three-nucleon
system, can be presented as a sum of three different Faddeev
components, each corresponding to different particle permutations and
projected on a particular set of partial waves \cite{Gloeckle}:
%
%
\begin{equation}
|\Psi \rangle = |\psi [(12)3]\rangle +
|\psi [(23)1]\rangle + |\psi [(31)2]\rangle .
\label{faddeev}
\end{equation}
The individual components read for instance in coordinate space
\begin{equation}
\psi^\nu (r_{ij},\rho_k)
= \langle r_{12}\rho_3\nu_{12} |\psi [(12)3]\rangle
= \langle r_{23}\rho_1\nu_{23} |\psi [(23)1]\rangle
= \langle r_{31}\rho_2\nu_{31} |\psi [(31)2]\rangle ,
\end{equation}
where $r$ and $\rho$ denote the pair and spectator coordinates,
respectively, and the index $\nu$ denotes the relevant quantum numbers.
In Ref. \cite{hgs} these Faddeev components were presented
in separable form $v^\nu(r_{ij})w^\nu(\rho_k)=\psi^\nu (r_{ij},\rho_k)$.

A parameterization of the individual Faddeev components is still rather clumsy
to use in situations where e.g. a two-body operator should act
on a term with a "wrong" grouping, i.e. the operator and the
pair wave function $v^\nu (r_{ij})$ do not involve the same particles.
However,  the action of a two-body operator on a completely antisymmetric
state does not depend on the particle identifications
in the operator. Therefore, we
parameterize the total antisymmetrized wave function
in terms of
pair and spectator coordinates (or momenta) projected on
different angular momentum eigenstates. We choose
the particle pair $(12)$ as the active pair and define
\begin{eqnarray}
\nonumber
v^\nu(r)w^\nu(\rho) &=& \langle r\, \rho \, \nu|\Psi\rangle \\
\nonumber
&=& \langle r \rho \nu  |\psi [(12)3]\rangle
+\sum_{\nu_{23}}\langle r\rho\nu|r_{23}\rho_{1}\nu_{23}\rangle
\langle r_{23}\rho_1\nu_{23} |\psi [(23)1]\rangle \\
& & \hskip 2.65cm
+\sum_{\nu_{31}}\langle r\rho\nu|r_{31}\rho_{2}\nu_{31}\rangle
\langle r_{31}\rho_2\nu_{31} |\psi
[(31)2]\rangle \ .
\label{def}
\end{eqnarray}
Here the $\langle r\rho\nu|r_{ij}\rho_{k}\nu_{ij}\rangle$ denote the
necessary recoupling coefficients for angular momentum, spin and isospin
and the $r_{ij}$ and $\rho_k$ are to be expressed in terms of $r$ and $\rho$
\cite{Gloeckle}.
E.g. one finds $\vec r_{23}=-\frac{1}{2}\vec r - \vec \rho$.

Physically one
might expect that, if in a bound three-body system either the
spectator or the pair is far off-shell, then it would be less
likely to find also the other far off-shell. In other words,
there should be some kind of correlation between the momenta
or the corresponding coordinates, which is not present in the
simple product. In short: there is no reason to expect the
triton wave function to be separable.
We shall allow for such a correlation by using a fit  with a sum of two
products. Over a large momentum range this seems to give
sufficiently accurate results.

In order to get converged results for the trinucleon binding
energy and the wave function, in principle, a fairly large
number of three-nucleon partial waves is needed \cite{Friar}. However,
most of those partial waves give only a small contribution to the
total wave function, cf., e.g., Ref. \cite{Schadow}.
Therefore, in the following treatment we shall
restrict ourselves to the most important states, i.e. those
where the $NN$ pairs are in the singlet spin state
$^1s_0$ and in the triplet states $^3s_1$ or $^3d_1$.
In this case the trinucleon wave function $\psi^\nu$ has only
five components whose quantum numbers are summarized in
Table \ref{Qnum}. In line with this restriction we also
include only those five components in the evaluation of the
total wave function $|\Psi\rangle$ according to Eq.~(\ref{def}).
The higher partial waves induced by the antisymmetrization have much
less weight. We should mention, however, that they still
contribute to the overall normalization \cite{Afnan}.
But, for convenience, the total wave functions employed in the present
work are normalized in such a way that the sum over the five considered
channels adds up to unity.

The procedure for the fit is the following. First the full Faddeev
wave function (using the CD Bonn \cite{bonn} and
Paris potential \cite{paris}) is obtained numerically in momentum space,
taking into account the permutation of the particles (c.f. Eq. (\ref{def})),
and then projected on the 5 considered partial waves (channels).
This gives the
"exact" wave functions $\Psi^{\nu}(p,q)$ to be fitted in each of the
5 channel considered ($^1s_0S$, $^3s_1S$, $^3d_1S$,
$^3s_1D$ and $^3d_1D$, cf. Table \ref{Qnum})
as functions of the momenta
$p$ and $q$ for the pair and the spectator, respectively.
 Each of these is approximated by a product
of functions $v^{\nu}_1(p)$ and $w^{\nu}_1(q)$ given by a
five-term expansion of Lorentz functions,
\begin{equation}
v^{\nu}_{\lambda}(p) = \sum_{n=1}^5
\frac{a^{\nu}_{n,\lambda}}{p^2+(m^{\nu}_{n,\lambda})^2}\,
,\qquad
w^{\nu}_{\lambda}(q) = \sum_{n=1}^5
\frac{b^{\nu}_{n,\lambda}}{q^2+(M^{\nu}_{n,\lambda})^2} \, .
\end{equation}
Here for each $\nu$ and $\lambda$ the conditions
\begin{equation}
\sum_{n=1}^5 a^{\nu}_{n,\lambda} = \sum_{n=1}^5 b^{\nu}_{n,\lambda} = 0
\label{asum}
\end{equation}
are required for convergence in momentum space and
to guarantee regularity at the origin in coordinate space
(for $D$ waves more conditions are needed \cite{hgs,Lacombe}).
We perform
a standard $\chi^2$ fit to the total antisymmetrized exact
(numerical) three-nucleon wave functions by minimizing
the function
\begin{equation}
\int_0^{\infty} dp\; dq\; p^2q^2 |\Psi^{\nu}(p,q)-v^{\nu}_1(p) w^{\nu}_1(q)|^2.
\end{equation}
Like Hajduk et al. \cite{hgs} we include the additional
weight factor $p^2q^2$, for this factor
emphasizes more the relevant momentum range, as factors $p^2$ and $q^2$ always
appear naturally in any matrix element as they originate from the
integration measure.
After the first fit, where only the term $v^{\nu}_1(p) w^{\nu}_1(q)$
is included, an additional
product term $v^{\nu}_2(p) w^{\nu}_2(q) $ is fitted to the remaining
deviation between this fit and the exact wave function.
Thus we have
\begin{equation}
\Psi^{\nu}(p,q)=v^{\nu}_1(p)w^{\nu}_1(q)+v^{\nu}_2(p)w^{\nu}_2(q).
\label{twoterms}
\end{equation}
In this work we stop after the second term. However, it should be clear
that every additional term included systematically improves the
fit without influencing the previous terms.

\begin{figure}[t!]
\begin{center}
\epsfig{file=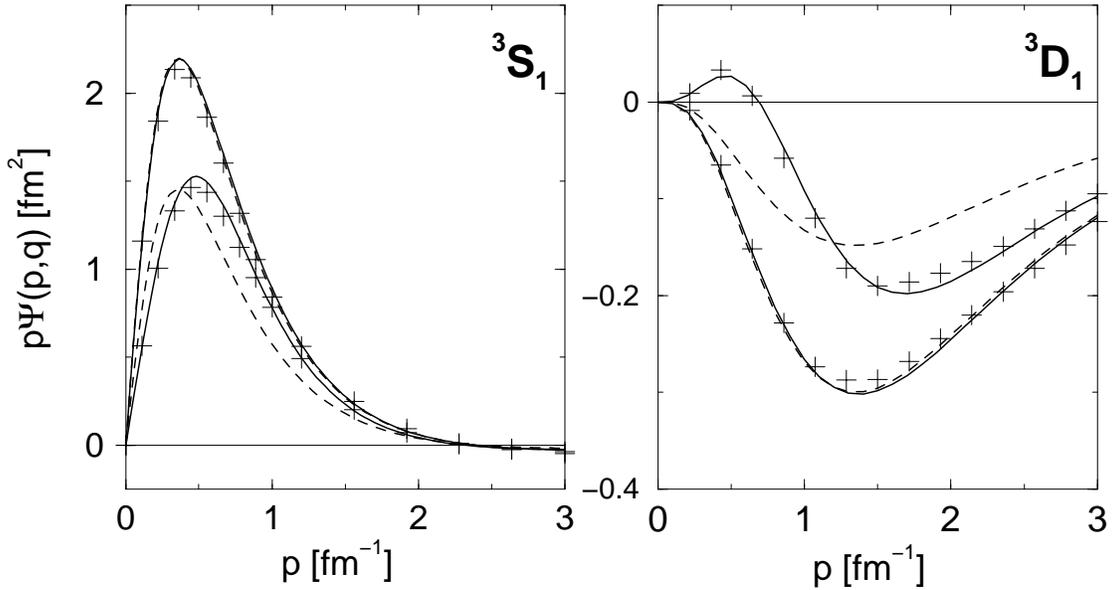, height=8cm, angle=0}
\caption{The full antisymmetric wave function
for the channels $^3s_1S$ (left panel) and $^3d_1S$ (right panel) as a function
of the relative pair momentum $p$ for spectator momenta $q= 0.5$
(larger pair) and 1 fm$^{-1}$ (smaller pair)---the latter multiplied
by a factor of three. For both momenta we show the
single term fit (dashed line), the two term fit (solid line) and the exact wave
function ($+$).
Here we used the wave function derived from the CD Bonn potential.}
\label{erg}
\end{center}
\end{figure}

We found that the inclusion of two terms is sufficient to reproduce
the "exact" wave function well over a reasonably large momentum range.
This is illustrated in Fig. \ref{erg}, where the one and two term
fits are compared to the "exact" wave function in the $^3s_1S$ as well
as the $^3d_1S$ channel for two different spectator momenta.
In particular for the $^3d_1S$ case the need for a second term is
striking.
Another way of testing the quality of the analytical
parameterization is offered by
the probability for the different components of the
wave function in the Blatt-Derrick representation \cite{Schadow}.
Corresponding results, for the CD Bonn as well as for the Paris
potential, are compiled in Table \ref{probab}.
Here the apparent non-separability of the $^3d_1S$ is reflected
in the $D$-state probability $P_D$. With a single-term
fit there is still a significant deviation of the value fitted for
the CD Bonn potential (Paris potential) 6.995\% (8.319\%) from
the exact one 7.127\% (8.428\%), while the two-term fit has
essentially converged to the exact result with a significantly
better value 7.082\% (8.386\%), cf Table \ref{probab}.
%

In the coordinate space representation these approximate wave
functions will consist of sums of Yukawa functions and
(for $D$ waves) their derivatives
\begin{eqnarray}
\nonumber
V^{\nu}_{\lambda}(r) &=& \sqrt{\frac \pi 2}\,\sum_{n=1}^5 a^{\nu}_{n,\lambda}\,
\, e^{-m^{\nu}_{n,\lambda}r}\quad
{\rm or}\\
V^{\nu}_{\lambda}(r) &=& \sqrt{\frac \pi 2}\,\sum_{n=1}^5 a^{\nu}_{n,\lambda}\,
\, e^{-m^{\nu}_{n,\lambda}r}
\left( 1+\frac{3}{m^{\nu}_{n,\lambda}r}+\frac{3}{(m^{\nu}_{n,\lambda}r)^2}\right)\, ,
\end{eqnarray}
with similar expressions for the spectator $\rho$ dependence.
As usual, in coordinate space an additional factor of $r$ is included
in the definition of the wave functions ($V$ is $r$ times the
Fourier transform of $v$).

Table \ref{tablebonn} contains the parameters for the fit to the wave
functions using the CD Bonn potential. It may be noted that
there is some freedom in distributing the strength between the
pair functions $v^{\nu}_{\lambda}$ and the spectator $w^{\nu}_{\lambda}$.
This freedom is used to present
the results in such a normalization that the square of
the first term pair function integrates to unity, i.e.
$\int_0^{\infty} dp\; p^2 (v_1^{\nu})^2=1$ for each partial wave $\nu$.
Table \ref{tableparis} gives the parameters of the triton
wave function fit for the Paris potential. The parameters
are available from the authors by e--mail\footnote{Either from
J. Niskanen (Jouni.Niskanen@helsinki.fi) or from J. Haidenbauer
(j.haidenbauer@fz-juelich.de).}.

As mentioned before, the parametrized wave functions $\Psi^{\nu}(p,q)$ are
normalized in the usual way, i.e.
\begin{equation}
\sum_{\nu}\int_0^{\infty} dp\; dq\; p^2q^2 |\Psi^{\nu}(p,q)|^2=1,
\label{norm}
\end{equation}
for the sum over the five channels considered. However, for completeness
reasons we would like to point out
that in a fully converged antisymmetrization of the wave function, with
sufficient higher partial wave included, those five
channels would have a relative weight of 0.954 (CD Bonn) and
0.945 (Paris), respectively.
We would like to stress that the remaining 5 \% are saturated by a large number
of partial waves that are individually rather small (c.f. Ref. \cite{Schadow})
and thus should be of minor relevance for the calculation of observables.

\begin{figure}[tb]
\begin{center}
\epsfig{file=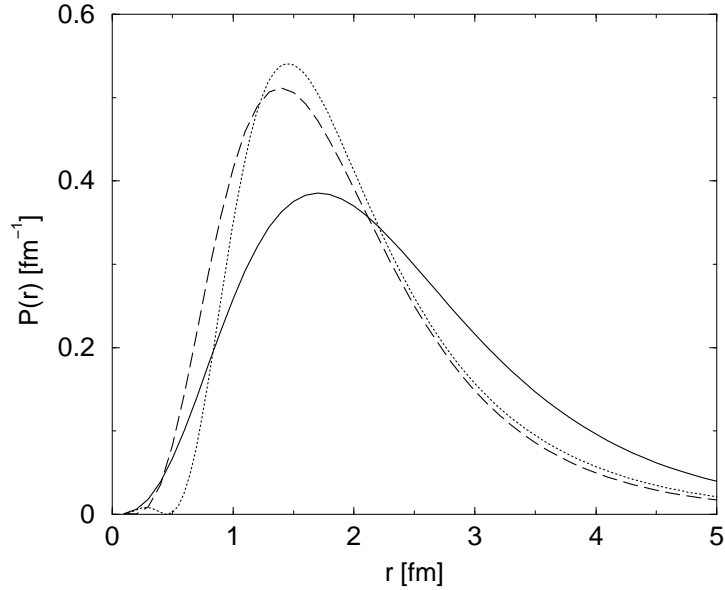,height=8cm, angle=0}
\caption{The coordinate space probability distributions for
the channel $^3s_1S$ with the spectator motion integrated out.
Dotted curve: the
representation of the Faddeev amplitude of Hajduk {\it et al.}
\protect\cite{hgs}; Dashed line: the same quantity, however, showing
our fit to the wave function derived from the Bonn CD potential (c.f.
table \protect\ref{tablesingle});
Solid line: full antisymmetrized wave function. }
\label{coordspwf}
\end{center}
\end{figure}

Fig.
\ref{coordspwf} shows the probability distributions
(pair correlations)
\begin{equation}
P(r)  \equiv \int_{0}^{\infty} |\Psi^{(^3s_1S)}(r,\rho)|^2 d\rho
\end{equation}
integrated over the spectator degrees of freedom. For
definiteness, these are all normalized to unity.
Here, for comparison with the earlier work \cite{hgs}
(dotted line), we also include the fits to the
individual Faddeev amplitude (dashed line). As one
might expect, these results are rather similar to
each other, while the antisymmetrized wave function
gives a significantly longer ranged distribution
(solid line). It is
interesting that the short-range node in the wave functions
of Ref. \cite{hgs} is also present in our single-channel fits
but not in the full antisymmetric wave function. For
completeness, Table \ref{tablesingle} gives the parameterization
fully analogous to Ref. \cite{hgs} used to produce the dashed
curve. With the exception of the dotted line, the results in this figure
are based on the  CD Bonn potential.

\section{The $\pi\, ^3$H\lowercase{e} scattering length}

Now we want to demonstrate the usefulness of the new parameterization by
an explicit calculation of the real part of the $\pi\, ^3$He scattering length in chiral
perturbation theory. Since
all the complications of having to treat a three-nucleon problem
were already solved on the level of deriving the wave function
parameterization in terms of an active pair and a spectator
(c.f. Eq. (\ref{def})),
the remaining part of the calculation for the scattering length is
as complicated as the corresponding calculation for the two nucleon
system (cf., e.g., Refs. \cite{Ulf,bk,Ericson} where the $\pi d$
scattering length was
calculated), as long as only two nucleon operators are
included as is the case for the NLO calculation we are
going to perform here. Note, however, that with the parameterization
given also the explicit evaluation of three body forces is
largely simplified. In addition, the very simple form of the terms in the separable
expansion allows an analytical calculation---the final result can
be expressed in terms of incomplete $\Gamma$-functions.

We will work with the so called hybrid approach, introduced by Weinberg
\cite{Weinberg}, namely the $\pi NNN \to \pi NNN$ transition operators,
evaluated using chiral perturbation theory, will be convoluted with
phenomenological wavefunctions using the parameterizations given in the
previous section. It should be stressed that recently three-body wavefunctions
were derived utilizing $NN$ and $3N$ forces that are
consistent with the chiral counting scheme \cite{Epelbaum1}.
As will become clear below, once a parameterization in the form presented
in this paper for those wavefunctions is available as well, it is a simple
task to extract the corresponding $\pi \, ^3$He scattering length.

\begin{figure}[t!]
\begin{center}
\vspace{3cm}
\epsfig{file=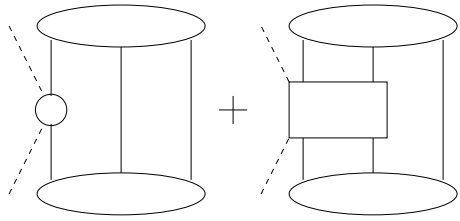, width=6cm}
\caption{Leading order contributions to the $\pi\, ^3$He scattering
length.}
\label{diafull}
\end{center}
\end{figure}

Within the counting scheme advocated by Weinberg all that contributes to
$\pi\, ^3$He scattering at leading and
next--to--leading order are one- and two-body currents, as illustrated
in Fig. \ref{diafull}. Due to the Goldstone nature of the
pions all three-body currents are suppressed by an additional
factor $p^2/\Lambda^2$, where $p$ denotes the typical momentum of
the problem and $\Lambda$ denotes the chiral symmetry breaking scale
($\simeq 1$ GeV)
\cite{Weinberg}.

In order to have a theory with predictive power it is compulsory that
unknown counter terms
 are either suppressed or at least small in number.
 Naturally,
in different schemes those terms appear at different places. This is not only true
for the three-body system itself, where in a pion--less approach three-body
forces appear at leading order \cite{paulo}, but also for the scattering
process. For example in Ref. \cite{harald} it was stressed, that in an approach
using perturbative pions there appear counterterms in the production
operator already
at leading order in a calculation for $\pi d$ scattering. However, in
this paper we will not elaborate on those subtleties any further but
will use the approach of Ref. \cite{Weinberg}.

It is a straightforward task to relate the
one body contribution to the scattering length. One gets
\begin{equation}
a^{(1b)}_{\rm He}= \kappa
\left( Aa^{(+)}-Q_\pi \, a^{(-)}2T_3 \right) \ ,
\label{finala1}
\end{equation}
where $\kappa = \left(1+\frac{m_\pi}{m_N}\right)/\left(1+\frac{m_\pi}{Am_N}\right)$.
Here $a^{(+)}$ ($a^{(-)}$) denote the isoscalar (isovector) $s$--wave $\pi N$
scattering length, $A$ is the number of nucleons in the nucleus (here
$A=3$), $Q_\pi$ is the charge of the pion in units of $e$ and $T_3$ denotes
the third component of the isospin of the nucleus.
Since the interaction is momentum independent, the loops occurring are
identical to the expression of the wave function normalization.
In contrast to the case of pion scattering on the deuteron, here
the isovector piece of the scattering length does contribute and dominates
the one body contribution. We thus get for the one body contribution of
the scattering of negatively charged pions on $\, ^3$He
\begin{equation}
a^{(1b)}_{\rm He} = (92 \pm 15)\times 10^{-3} \ m_\pi^{-1} \ .
\label{1b}
\end{equation}
where we used $a^{(-)}=(90.5\pm 4.2)\times 10^{-3} \ m_\pi^{-1}$ and
$a^{(+)}=(-2.2\pm 4.3)\times 10^{-3} \ m_\pi^{-1}$.
These are the purely hadronic values presented in Ref. \cite{schr}, where
Coulomb effects as well as isospin breaking effects are already subtracted.
Note, that a consistent next--to--leading order calculation would require
to use $\pi N$ scattering lengths extracted from a calculation for
$\pi N$ scattering carried
out to the corresponding order as given, e.g., in Ref. \cite{nadia}.
In addition, a careful analysis would also
necessitate to take into account isospin breaking effects consistently,
as was stressed in Ref. \cite{silas}.
One expects, however, that especially the latter point enlarges primarily
the theoretical error.
Anyway, here we will ignore such subtleties because we only want 
to investigate, more qualitatively, if the $\pi \, ^3$He system
allows to extract information on the $\pi N$ scattering lengths.

Now we need to evaluate the matrix element of  the wave function
with the two-body scattering kernel
${\cal A}(\vec p \, ',\vec p)$, where $\vec p$ ($\vec p \, '$) denote
the relative momentum of the active nucleon pair in the initial (final)
state, respectively.
Thus, we can write
\begin{equation}
a^{(2b)}_{\rm He}=\frac{1}{4\pi}{A \choose
2}\left(\frac{1}{1+\frac{m_\pi}{Am_N}}\right)
\int d^3p'd^3  p d^3 q\Psi (\vec p \,', \vec q)^\dagger
{\cal A}(\vec p \, ',\vec p)\Psi (\vec p, \vec q) \ .
\label{a2b}
\end{equation}
Here we already used the property that the wave function is antisymmetrized.
Thus, whichever of
the three nucleons is the spectator, the result of the matrix element
is the same. Therefore, inclusion of all possible pairwise
interactions just leads to an overall factor of 3. This factor can
be easily generalized to the case of a nucleus composed of $A$ nucleons.
Then there is a combinatorial factor which
is given by the binomial coefficient ${A \choose 2}$. If we now use the
separable form presented in the previous section, we get
\begin{equation}
a^{(2b)}_{\rm He}=\frac{1}{4\pi}{A \choose 2}
\left(\frac{1}{1+\frac{m_\pi}{Am_N}}\right)
\sum_{\nu \nu '}\sum_{\lambda ' \lambda}{\cal W}^{\nu' \nu}_{\lambda ' \lambda}
\int d^3p'\, d^3  p \, v (p \, ')_{\lambda '}^{\nu '}
\langle \nu ' ,\hat p \, '|{\cal A}(\vec p \, ',\vec p)| \nu , \hat p \rangle v (
p)_\lambda^{\nu} \ ,
\end{equation}
where
\begin{equation}
{\cal W}^{\nu ' \nu}_{\lambda ' \lambda} = \int d^3 q\, w(q)_{\lambda '}^{\nu '}
w(q)_\lambda^{\nu}
\langle \nu ' ,\hat q| \nu , \hat q \rangle \ .
\end{equation}
Here $\lambda$ and $\lambda '$ denote the terms of the separable ansatz,
$\nu '$ as well as $\nu$ denote the partial waves for the final
and initial state, respectively,
and the different state vectors $|\nu, \hat p \rangle$
contain the relevant spin, isospin
and angular momentum components of the corresponding partial wave.

\begin{figure}[t!]
\begin{center}
\vspace{1cm}
\epsfig{file=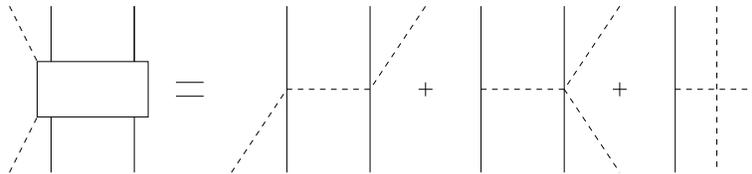, width=10cm}
\caption{Leading two-body contributions to the $\pi\, ^3$He scattering
length.}
\label{dia}
\end{center}
\end{figure}

For the two body scattering kernel, as shown diagrammatically in Fig.
 \ref{dia}, we use \cite{Ulf}:
\begin{eqnarray}
\nonumber
{\cal A}(\vec p \, ',\vec p)
 &=& \frac{1}{(2\pi)^3}\frac{m_\pi^2}{4F_\pi^4}
 \left\{ \frac{1}{\vec q {}\, ^2}\,
[2\delta^{(ac)}(\vec \tau ^{(1)}\cdot \vec \tau^{(2)})-
\tau ^ {(1)  a}\tau ^{(2)  c}
-  \tau ^ {(1)  c}\tau ^{(2)  a}]\right. \\
&-& \left. g_A^2\frac{(\vec \sigma ^{(1)} \cdot \vec q)
(\vec \sigma ^{(2)} \cdot \vec q)}{(\vec q {}\, ^2+m_\pi^2)^2}
\, [\delta^{(ac)}(\vec \tau ^{(1)}\cdot \vec \tau^{(2)})-
 \tau ^ {(1)  a}\tau ^{(2)  c}
-  \tau ^ {(1)  c}\tau ^{(2)  a}]\right\} \ ,
\end{eqnarray}
where the superscripts $a$ ($c$) denote the isospin of the incoming (outgoing) pion,
and $\vec q = \vec p\, ' - \vec p$ denotes the pion momentum.
It should be stressed that the same kernel was used recently to extract
the isoscalar $\pi N$ scattering length from $\pi d$ scattering data \cite{Ulf}.
Thus---as long as we restrict ourselves to $S$--waves in the bound
state wave functions---all we need to evaluate are the following two
integrals
\begin{eqnarray}
I_1(m_i,m_j) &=& \frac{1}{4\pi} \int d^3p\, d^3p'\,
\frac{1}{\vec p {}\, ^2+m_i^2}\,\frac{1}{(\vec p-\vec p \, ')^2}\,
\frac{1}{\vec p \, ' \, ^2 +m_j^2} \\
I_2(m_i,m_j) &=& \frac{1}{4\pi} \int d^3p\, d^3p'\,
\frac{1}{\vec p {}\, ^2+m_i^2}
\frac{(\vec p-\vec p \, ')^2}{((\vec p-\vec p \, ')^2+m_\pi^2)^2}\,
\frac{1}{\vec p \, ' \, ^2 +m_j^2} \ .
\end{eqnarray}
Obviously the inclusion of $D$ waves can be done easily. Since the
formulas are rather lengthy we do not give them here explicitly. It
should be stressed that
the contribution due to the $D$ waves did not exceed 10\%, which is of
the order of the theoretical uncertainty induced by both
the convergence rate of the chiral expansion and the model
dependence of the scattering length caused by the particular employed
nuclear wave function. We find
\begin{eqnarray}
I_1(m_i,m_j) &=& \pi^3\Gamma ( (m_i+m_j)\epsilon,0)
\label{int1}
\\
I_2(m_i,m_j) &=& \pi^3\left\{ \Gamma
((m_i+m_j+m_\pi)\epsilon,0)
-\frac{1}{2}\left(\frac{m_\pi}{m_i+m_j+m_\pi}\right) \right\} \ ,
\label{int2}
\end{eqnarray}
where we used the incomplete $\Gamma$-function defined through
$\Gamma (\lambda , n)
=\int_\lambda^\infty dx\, x^{n-1}\exp (-x)$ \cite{aseg}.
The constant $\epsilon$ was introduced to render the integrals finite.
Eq. (\ref{asum}) implies that the results in
Eqs. (\ref{int1}) and (\ref{int2}) are independent of $\epsilon$ as long as
$m_i \epsilon \ll 1$ for all $i$.
%
%
%

We thus get for the two-body contribution to the $\pi \, ^3$He scattering
length
\begin{eqnarray}
\nonumber
a^{(2b)}_{\rm He} &=&{A \choose 2}\left(\frac{1}{1+\frac{m_\pi}{Am_N}}\right)
\left(\frac{m_\pi^2}{32\pi^4F_\pi^4}\right)
\sum_{\nu }\sum_{\lambda ', \lambda=1}^2{\cal W}^{\nu \nu}_{\lambda ' \lambda}
\left(\frac{2}{3}T_{\nu}\left(T_{\nu}+1\right)-1\right)  \\
& \times & \sum_{m,n=1}^{5} a_{m,\lambda '}^{\nu}a_{n,\lambda}^{\nu}
\left\{I_1(m_{m,\lambda '}^{\nu}
,m_{n, \lambda}^{\nu})
-\frac{g_A^2}{4}\left(\frac{2}{3}S_{\nu}\left(S_{\nu}+1\right)-1\right)
I_2(m_{m,\lambda '}^{\nu},m_{n,\lambda}^{\nu})\right\} .
\label{finala2}
\end{eqnarray}
Again, only the formula for the $S$--wave part of the wavefunction
is given explicitly, although the effect of the $D$--waves is
included in the final result. Under this restriction, the partial
wave of the initial $NN$ pair is equal to that of the outgoing pair.

Using the parameters for the two different wave functions as listed
in Tables \ref{tablebonn} and \ref{tableparis},
we get for the two-body contributions to the real part of the scattering length
\begin{equation}
a^{(2b)}_{\rm He} = \left\{
\begin{array}{l} -24\times 10^{-3}\; m_\pi^{-1} \ {\rm (Paris)} \\
 -26\times 10^{-3}\; m_\pi^{-1} \ {\rm (CD\; Bonn)} \ . \end{array} \right.
 \end{equation}
Compared to the leading one-body term we thus find reasonable convergence.
Also the dependence on the wave function used can be regarded as a
higher order effect.
As the final result we therefore have
\begin{equation}
{\rm Re}(a_{\rm He}^{\rm (theo)}) = (67 \pm 15)\times 10^{-3} \ m_\pi^{-1} \ .
\label{theo}
\end{equation}
The error included is only that stemming from the error on the
$\pi N$ scattering lengths used (c.f. Eq. (\ref{1b})). As mentioned
before, for a reliable estimate of the theoretical error a
consistent calculation including isospin breaking effects as well
as using consistent wave functions would
be necessary. One could regard the size of the NLO contribution as
a rather conservative estimate for the theoretical error.

Information on the $\pi \, ^3$He scattering length
can be inferred from experimental information on level shifts caused
by the strong interaction in the bound state energies of
$\pi^- {}\, ^3$He atoms \cite{phe1,phe2,phe3}.
A list of results from different experiments available
in the literature is given in table \ref{scltab}. From those
energy shifts scattering lengths
can be easily extracted using the so called Deser formula \cite{deser},
\begin{equation}
\epsilon_{1s}=-2(Z \alpha)^3\mu^2\, {\rm Re}\left( a_{\rm He}^{\rm (exp)} \right) \ .
\label{scl}
\end{equation}
where $Z$ is the charge number of the nucleus and $\mu$ is the reduced
mass of the $\pi\, ^3$He system.
Taking the arithmetic mean value for the experimental results
given in table \ref{scltab}, we thus get
\begin{equation}
{\rm Re}(a_{\rm He}^{\rm (exp)}) = (47 \pm 13)\times 10^{-3} \ m_\pi^{-1} \ ,
\label{exp}
\end{equation}
where the error includes both the spread in the experimental numbers
as well as an additional 10\% to account for the omission of
electromagnetic and isospin breaking corrections in Eq. (\ref{scl}).
Note that in case of pionic hydrogen those corrections turned out to be of
the order of 7\% \cite{gasser}. Thus, we find agreement between 
theory and experiment within the given uncertainties.

As was argued above, at the current stage we cannot give a reliable
error estimate for the theoretical calculation.
Naturally, further studies are needed to
draw stronger conclusions. However, it should be stressed, that
it would be very useful to push the calculation as well as the
experiment for the $\pi \, ^3$He system to as high an accuracy as those for the
$\pi$d system, since remaining discrepancies between theory and
experiment in such a combined analysis can only stem from three-body
currents and would therefore be an important test of chiral perturbation
theory in few body systems.

In this context let us also emphasize
that an experimental study of pionic tritium would
be of very high interest because a combined analysis of the $\pi \, ^3$He
and the $\pi t$ systems promises to provide direct access to the
pion--nucleon scattering lengths. From Eqs. (\ref{finala1}) and (\ref{finala2}) we
find
\begin{eqnarray}
a_t+ a_{\rm He} &=&
6\kappa a^{(+)}+{\cal O}\left(\left(\frac{m_\pi}{M_N}\right)^3\right) \ ,
\nonumber \\
a_t- a_{\rm He} &=&
2\kappa Q_\pi \, a^{(-)} + {\cal
O}\left(\left(\frac{m_\pi}{M_N}\right)^4\right)
 \ ,
\label{asad}
\end{eqnarray}
where the constant $\kappa$ was defined in Eq. (\ref{finala1}).
Note, that for the extraction of $a^{(-)}$ one should expect the
corrections to be suppressed by one chiral order compared to that
of $a^{(+)}$. This is the case, because the leading corrections, 
that in the counting advocated by Weinberg 
appear at ${\cal O}((m_\pi /M_N)^3)$---as given
in Eq. (\ref{finala2}) for only $s$--waves in the bound state wavefunction---do
 not distinguish between the $t$ and $^3$He system
and thus do not contribute to  the difference of the corresponding scattering lengths. 
Indeed a study of $\pi^- t$ bound states appears
experimentally feasible nowadays \cite{gotta}. It should be stressed,
that Eqs. (\ref{asad}) are general and independent of the wave functions used.
The error estimate only requires that the Weinberg scheme is applicable to the
reactions under investigation. In addition, it should be clear
that the $^3$He/$t$ system is an ideal system to study isospin breaking effects
in few nucleon systems.

It turns out that for the $\pi\, ^3$He scattering
the omission of the second term in the expansion of the wave function
(c.f. Eq. (\ref{twoterms})) leads to a change
in the two-body contribution of the scattering length by 2\% only.
Thus, in the energy range relevant
for $\pi$ scattering at threshold, the full wave function is well
described by a one-term separable form.
As indicated in Fig. \ref{erg} in reactions with either large momentum
transfer or where higher partial waves become relevant, the effect
of the second term is expected to be much more pronounced.
Indeed, this has been already seen in a recent application of the
wave function parametrization to $\pi$ absorption
on $^3$He \cite{piabs}.

It is illuminating to compare Eq. (\ref{finala2}) with that for
the two-body contribution in the case of pion-deuteron scattering.
Obviously, in this situation it does not make sense to use two different
terms expanded individually (thus the summation on $\lambda$ and
$\lambda '$ disappears). In addition, there is no third particle (thus
${\cal W}^{\nu \nu}_{\lambda ' \lambda}=1$). All the rest remains
unchanged. If we now use for the deuteron wave function a
parametrization that was provided in Ref. \cite{Machleidt}
for the Bonn B $NN$ model we reproduce the
result obtained in Ref. \cite{Ulf}.

In the present work we do not aim at a highly accurate determination
of the $\pi\, ^3$He scattering length, but rather at a demonstration of
the usefulness of the new parameterization. However, it should be
clear that it is straightforward to improve the calculation. Results
to order $p^4$ for the $\pi d$ scattering length are given in Ref.
\cite{silas} and could be used as the basis for an equally accurate
calculation for the $\pi \, ^3$He scattering length.

Already in Ref. \cite{Ulf} it was observed, that the contribution from
the first diagram on the right hand side in Fig. \ref{dia} is
significantly larger than that from the other two, although all three
formally appear at the same order in the Weinberg scheme. In Ref.
\cite{silas} this was traced back to the small momentum scale
introduced by the small binding energy of the deuteron. To account for
this a modified counting scheme was advocated, leading to a reordering
of the meson exchange diagrams: the first diagram on the right hand
side of Fig. \ref{dia} is now the leading two nucleon current, whereas
the other two are down by 4 orders in the expansion parameter (c.f.
table 4 in Ref. \cite{silas}).  It is interesting to note, that
in our calculation we find a similar hierarchy of the results for
the diagrams shown in Fig. \ref{dia}: the result
for the first diagram exceeds the sum of the latter two by a factor of
20 or more, depending on the wavefunction used.  Note that, similarly
to the case of the deuteron, the typical momentum of a nucleon within
$\, ^3$He is significantly smaller than the pion mass.

Also a calculation of the imaginary part of the scattering
length is, in principle, feasible. However, the contributions to the
imaginary part are connected
to the pion absorption on $^3$He, where a different counting
scheme needs to be applied to account for the large nucleon momenta
in the purely nucleonic intermediate state \cite{pwaves}.
However, within this scheme the imaginary part should be suppressed
by a factor of $(m_\pi/M_N)^{(3/2)}$ compared to Re($a^{2b}$)
given in Eq. (\ref{finala2}). The
same holds true for the so called dispersive corrections, namely the
contribution of the inelastic channels to the real part of the scattering length.

\section{Summary}

In this paper we have given a new parameterization for
two different trinucleon wave functions based on modern
$NN$ interaction models. Although the procedure
is in spirit similar to an earlier work \cite{hgs},
it extends the single-term separability to two terms
allowing some correlation between the two relative
momenta (or coordinates) and also approximates directly the
full antisymmetrized wave function projected onto different
partial waves. The former feature might be expected to
be useful in phenomena involving high momenta, whereas the
latter drastically simplifies calculations involving
two-nucleon operators only.

Apparently similar benefits are expected in
applications to other reactions.
E.g., a calculation of quasifree
pion absorption on nucleon pairs in $^3$He based on the
presented parameterization has been reported in Ref. \cite{piabs}.

As a demonstration
the parameterization has been applied to calculate the $\pi\, ^3$He
scattering length. It has been seen that for such a parameterization
the two-body contribution becomes very easy to compute even
analytically. The actual calculation done in NLO chiral
perturbation theory yields a value of
${\rm Re}(a_{\rm He}) =
(67 \pm 15)\times 10^{-3} \ m_\pi^{-1}$, which is in
reasonable agreement with experimental values
inferred from data on level shifts in pionic $^3$He bound states.
We have argued, that a combined analysis of $\pi d$, $\pi \, ^3$He and
$\pi t$ scattering should provide both  important information on
the $\pi N$ $s$--wave scattering lengths as well as
a test of the applicability of chiral perturbation theory
to few-body systems.

\acknowledgements{
We would like to thank D. Gotta for providing valuable information
on experiments on pionic atoms. Furthermore, we acknowledge
useful discussions with I.R. Afnan, H.W. Griesshammer, U.--G. Mei\ss ner,
and J. Speth.
Financial support for this work was provided in part by the
international exchange program between DAAD (Germany, project no.
313-SF-PPP-pz) and the Academy of Finland (project no. 41926) as well as
 by RFBR (grant no. 02-02-16465).}

\newpage

\begin{table}[ht]
\caption{Quantum numbers of the three-body channels.
$s$, $\tau$, $l$, and $j$ refer to the spin, isospin, orbital
and total angular momentum in the $NN$ subsystem and $L$ and $K$
are the relative orbital angular momentum of the spectator and
the so-called channel spin \protect\cite{Schadow}.}
\label{Qnum}
\begin{center}
\begin{tabular}{ccccccccc}
 Channel no. & Label & Subsystem & $l$ & $s$ & $j^\pi$ & $\tau$ & $K$ & $L$ \\
\hline
 1 &$^1s_0S$& $^1s_0$ & $0$ & $0$ & $0^+$ & $1$ & $1/2$ & $0$ \\
 2 &$^3s_1S$& $^3s_1$ & $0$ & $1$ & $1^+$ & $0$ & $1/2$ & $0$ \\
 3 &$^3s_1D$& $^3s_1$ & $0$ & $1$ & $1^+$ & $0$ & $3/2$ & $2$ \\
 4 &$^3d_1S$& $^3d_1$ & $2$ & $1$ & $1^+$ & $0$ & $1/2$ & $0$ \\
 5 &$^3d_1D$& $^3d_1$ & $2$ & $1$ & $1^+$ & $0$ & $3/2$ & $2$ \\
\end{tabular} \end{center}
\end{table}

\begin{table}[ht]
\caption{Probabilities (in the Blatt-Derrick representation) of
the trinucleon wave function components for the CD Bonn and
Paris potentials. ``1 term'' and ``2 terms'' refer to the single- and
two-term parameterizations described in Sect. II.
}
\label{probab}
\begin{center}
\begin{tabular}{lcccc} & P(S) & P(S') & P(P) & P(D) \\
\hline
\hline
 CD Bonn 1 term & 91.77 & 1.187 & 0.047 & 6.995 \\
 CD Bonn 2 terms & 91.60 & 1.274 & 0.048 & 7.082 \\
\hline
 CD Bonn & 91.55 & 1.276 & 0.049 & 7.127 \\
\hline
\hline
 Paris 1 term & 90.08 & 1.543 & 0.062 & 8.319 \\
 Paris 2 terms & 89.85 & 1.660 & 0.100 & 8.386 \\
\hline
 Paris & 89.90 & 1.610 & 0.066 & 8.428 \\
\end{tabular}
\end{center}
\end{table}

\begin{table}[tb]
\caption{The parameters $a_n$, $b_n$, $m_n$ and $M_n$ of the fit
for the full antisymmetric three-body wave function with the
CD Bonn potential.}
\label{tablebonn}
\begin{tabular}{lrrrr}
State/term & $a_n \ [fm^{-\frac{1}{2}}]$ & $m_n \ [fm^{-1}]$ & $b_n \
[fm^{-\frac{1}{2}}]
$ & $M_n \ [fm^{-1}]$ \\
\hline  \\
$^1s_0S$ &   -1.420402 &  0.408491 &   -3.269355 &  0.587360\\
Term 1  &   -0.713402 &  0.660926 &   11.674343 &  0.787360\\
 &   -1.500957 &  0.913361 &  -33.930463 &  0.987360\\
 &   10.077357 &  1.165797 &   44.864561 &  1.187360\\
 &   -6.442596 &  1.418232 &  -19.339085 &  1.387360\\
 \hline
Term 2  &    0.621726 &  0.346344 &    0.230330 &  0.384846\\
 &    8.073312 &  0.546344 &    5.555461 &  0.735757\\
 &  -39.510606 &  0.746344 &  -27.949487 &  1.086668\\
 &   52.419034 &  0.946344 &   38.650601 &  1.437579\\
 &  -21.603467 &  1.146344 &  -16.486905 &  1.788490\\
 \hline
$^3s_1S$  &    1.308614 &  0.424997 &   -4.354797 &  0.534346\\
Term 1  &    1.057018 &  0.818066 &   17.513718 &  0.734346\\
 &   -2.491967 &  1.211135 &  -42.954883 &  0.934346\\
 &   -3.044193 &  1.604204 &   49.777813 &  1.134346\\
 &    3.170527 &  1.997273 &  -19.981851 &  1.334346\\
 \hline
Term 2  &    0.668573 &  0.359204 &   -0.141018 &  0.319402\\
 &    8.171864 &  0.559204 &   -2.573848 &  0.702473\\
 &  -39.501304 &  0.759204 &   14.710142 &  1.085544\\
 &   51.788425 &  0.959204 &  -21.510088 &  1.468615\\
 &  -21.127558 &  1.159204 &    9.514812 &  1.851686\\
 \hline
$^3d_1S$ &    0.439080 &  0.550025 &   -0.260857 &  0.431744\\
Term 1  &    2.063638 &  1.214291 &    0.781535 &  0.705959\\
 &  -26.827630 &  1.878557 &   -4.200276 &  0.980175\\
 &   41.548562 &  2.542822 &    7.620099 &  1.254391\\
 &  -17.223651 &  3.207088 &   -3.940502 &  1.528606\\
 \hline
 Term 2 &   -0.068267 &  0.556661 &    0.458816 &  0.307065\\
 &    0.921148 &  1.096208 &   18.886372 &  0.828691\\
 &   -2.720463 &  1.635755 & -114.616019 &  1.350317\\
 &    2.869650 &  2.175302 &  176.411566 &  1.871943\\
 &   -1.002067 &  2.714849 &  -81.140735 &  2.393569\\
 \hline
 $^3s_1D$ &    7.856844 &  1.077420 &    0.008223 &  0.316062\\
Term 1  &   16.344288 &  1.277420 &    0.376375 &  0.820171\\
 &  -27.636271 &  1.477420 &   -2.558819 &  1.324281\\
 &  -26.578179 &  1.677420 &    3.621148 &  1.828390\\
 &   30.013317 &  1.877420 &   -1.446927 &  2.332499\\
 \hline
 Term 2 &   -1.089313 &  0.666635 &   28.192605 &  0.828086\\
 &    9.975483 &  0.866635 & -226.000462 &  1.203061\\
 &  -26.331496 &  1.066635 &  551.617505 &  1.578036\\
 &   27.163776 &  1.266635 & -530.511234 &  1.953012\\
 &   -9.718449 &  1.466635 &  176.701587 &  2.327987\\
 \hline
$^3d_1D$  &   -0.430426 &  0.345360 &    0.000745 &  0.180605\\
 Term 1 &   14.525808 &  0.921071 &   -0.016356 &  1.497104\\
 &  -59.378637 &  1.496783 &   -0.340857 &  2.813602\\
 &   72.417582 &  2.072494 &    0.621826 &  4.130100\\
 &  -27.134327 &  2.648206 &   -0.265358 &  5.446598\\
 \hline
 Term 2 &    0.017189 &  0.382127 &    1.692045 &  0.222982\\
 &   -0.110339 &  0.582127 &  -33.002368 &  0.422982\\
 &    0.209799 &  0.782127 &  119.183936 &  0.622982\\
 &   -0.160679 &  0.982127 & -139.556966 &  0.822982\\
 &    0.044030 &  1.182127 &   51.683353 &  1.022982\\

\end{tabular}
\end{table}

\begin{table}[tb]
\caption{The parameters $a_n$, $b_n$, $m_n$ and $M_n$ of the fit
for the full antisymmetric three-body wave function using the
Paris potential.}
\label{tableparis}
\begin{tabular}{lrrrr}
State/term & $a_n \ [fm^{-\frac{1}{2}}]$ & $m_n \ [fm^{-1}]$ & $b_n \
[fm^{-\frac{1}{2}}]
$ & $M_n \ [fm^{-1}]$ \\
\hline  \\
 $^1s_0S$ &    1.185873 &  0.366138 &    2.492877 &  0.537309\\
 Term 1  &    1.764563 &  0.647808 &   -8.176120 &  0.737309\\
 &   -3.192238 &  0.929479 &   26.193477 &  0.937309\\
 &   -3.305482 &  1.211149 &  -37.021445 &  1.137309\\
 &    3.547284 &  1.492820 &   16.511211 &  1.337309\\
\hline
  Term 2 &    0.071144 &  0.246611 &    0.130706 &  0.356517\\
 &   10.443707 &  0.455327 &    3.775052 &  0.687907\\
 &  -45.143466 &  0.664043 &  -18.874600 &  1.019297\\
 &   58.923642 &  0.872758 &   26.307399 &  1.350687\\
 &  -24.295028 &  1.081474 &  -11.338557 &  1.682077\\
\hline
 $^3s_1S$ &    1.129463 &  0.382407 &   -3.468556 &  0.482186\\
 Term 1  &    1.063093 &  0.803847 &   13.568797 &  0.682186\\
 &   -1.858398 &  1.225288 &  -34.549146 &  0.882186\\
 &   -4.412162 &  1.646729 &   41.474326 &  1.082186\\
 &    4.078003 &  2.068170 &  -17.025421 &  1.282186\\
\hline
 Term 2  &    0.536411 &  0.308809 &   -0.084904 &  0.285273\\
 &    9.613061 &  0.508809 &   -1.853015 &  0.653089\\
 &  -45.528006 &  0.708809 &   10.592513 &  1.020905\\
 &   60.473587 &  0.908809 &  -15.691503 &  1.388720\\
 &  -25.095054 &  1.108809 &    7.036910 &  1.756536\\
\hline
 $^3d_1S$ &    0.400574 &  0.519557 &   -0.238406 &  0.392032\\
  Term 1 &   -0.136721 &  1.295772 &    0.469922 &  0.693166\\
 &  -18.270490 &  2.071987 &   -3.110884 &  0.994299\\
 &   31.444828 &  2.848202 &    6.387897 &  1.295433\\
 &  -13.438191 &  3.624417 &   -3.508528 &  1.596567\\
\hline
 Term 2  &   -0.050950 &  0.504112 &    0.727812 &  0.331487\\
 &    0.788400 &  1.071790 &   17.453956 &  0.846965\\
 &   -2.417995 &  1.639468 & -115.129833 &  1.362444\\
 &    2.592890 &  2.207146 &  182.257098 &  1.877922\\
 &   -0.912345 &  2.774824 &  -85.309033 &  2.393400\\
\hline
 $^3s_1D$ &   -0.328513 &  0.432149 &    0.009219 &  0.297932\\
 Term 1  &    4.302698 &  0.813996 &    0.321081 &  0.804086\\
 &    7.984795 &  1.195843 &   -2.366748 &  1.310240\\
 &  -29.173703 &  1.577690 &    3.403634 &  1.816394\\
 &   17.214723 &  1.959537 &   -1.367186 &  2.322548\\
\hline
 Term 2  &   -0.028375 &  0.477081 &   53.452223 &  0.722757\\
 &    0.521663 &  0.777081 & -512.515615 &  1.152683\\
 &   -1.805123 &  1.077081 & 1359.198606 &  1.582609\\
 &    2.190226 &  1.377081 &-1367.181577 &  2.012535\\
 &   -0.878391 &  1.677081 &  467.046364 &  2.442462\\
\hline
 $^3d_1D$ &    0.053659 &  0.204481 &    0.000161 &  0.273758\\
 Term 1  &   -6.461875 &  0.882230 &   -0.187203 &  1.240855\\
 &   33.853359 &  1.559978 &    1.208634 &  2.207951\\
 &  -44.645453 &  2.237726 &   -1.685015 &  3.175047\\
 &   17.200310 &  2.915475 &    0.663424 &  4.142144\\
\hline
 Term 2  &   -0.065878 &  0.165941 &    0.300000 &  0.638723\\
 &    4.577657 &  0.797478 &   -2.649250 &  1.038723\\
 &  -17.610419 &  1.429016 &    6.528825 &  1.438723\\
 &   20.615853 &  2.060553 &   -6.235132 &  1.838723\\
 &   -7.517212 &  2.692091 &    2.055558 &  2.238723\\

\end{tabular}
\end{table}

\begin{table}[tb]
\caption{The parameters $a_n$, $b_n$, $m_n$ and $M_n$ of the fit
for single permutation Faddeev amplitudes for the CD Bonn
potential. Here the single permutation normalization integral defined
in Ref. \protect\cite{hgs} is $N^2 = \langle\psi(1)|\psi(1)\rangle
= 0.1597$. }
\label{tablesingle}
\begin{tabular}{lrrrr}
State/term & $a_n \ [fm^{-\frac{1}{2}}]$ & $m_n \ [fm^{-1}]$ & $b_n \
[fm^{-\frac{1}{2}}]
$ & $M_n \ [fm^{-1}]$ \\
\hline  \\
$^1s_0S$ &   -1.688022 &  0.532323 &   -0.455000 &  0.499291\\
Term 1 &    3.387087 &  0.945460 &   -0.018469 &  0.699291\\
 &  -19.727502 &  1.358596 &   -0.116043 &  0.899291\\
 &   38.728212 &  1.771733 &    1.839667 &  1.099291\\
 &  -20.699775 &  2.184870 &   -1.250155 &  1.299291\\
   \hline
Term 2  &    0.085356 &  0.265782 &   -0.000139 &  0.107800\\
 &    7.248584 &  0.676027 &    0.331163 &  0.461937\\
 &  -38.887127 &  1.086272 &   -1.640077 &  0.816075\\
 &   58.934705 &  1.496517 &    2.272345 &  1.170212\\
 &  -27.381518 &  1.906762 &   -0.963293 &  1.524350\\
      \hline
$^3s_1S$  &   -1.426869 &  0.510733 &    0.003479 &  0.175113\\
Term 1  &    0.526004 &  0.973756 &    0.819387 &  0.436154\\
 &   -6.707522 &  1.436779 &   -0.201750 &  0.697195\\
 &   18.647993 &  1.899802 &   -1.766824 &  0.958236\\
 &  -11.039606 &  2.362825 &    1.145708 &  1.219277\\
      \hline
Term 2  &    0.563686 &  0.311171 &   -0.111000 &  0.341604\\
 &    5.367244 &  0.753186 &   -0.665672 &  0.567744\\
 &  -39.146731 &  1.195201 &    4.131617 &  0.793884\\
 &   63.040222 &  1.637216 &   -5.800799 &  1.020025\\
 &  -29.824421 &  2.079230 &    2.445854 &  1.246165\\
      \hline
$^3d_1S$ &    0.278338 &  0.476215 &   -0.191666 &  0.416574\\
Term 1  &    1.922824 &  1.189219 &    0.132050 &  0.667231\\
 &  -23.998840 &  1.902224 &   -1.267253 &  0.917888\\
 &   37.142550 &  2.615229 &    2.720458 &  1.168546\\
 &  -15.344872 &  3.328234 &   -1.393589 &  1.419203\\
     \hline
Term 2  &   -0.008422 &  0.434522 &    2.864122 &  0.388367\\
 &    0.192188 &  1.123081 &   16.511009 &  0.717339\\
 &   -0.653916 &  1.811640 & -116.582783 &  1.046312\\
 &    0.733997 &  2.500199 &  174.361137 &  1.375285\\
 &   -0.263848 &  3.188758 &  -77.153484 &  1.704258\\
      \hline
$^3s_1D$  &    0.761443 &  0.575896 &    0.020358 &  0.389875\\
Term 1  &    1.462244 &  0.775896 &   -0.171640 &  0.897031\\
 &   -0.354613 &  0.975896 &    0.119029 &  1.404187\\
 &    1.482841 &  1.175896 &    0.131303 &  1.911344\\
 &   -3.351914 &  1.375896 &   -0.099050 &  2.418500\\
    \hline
Term 2  &   -0.010448 &  0.610872 &    3.527742 &  0.888433\\
 &   -0.194635 &  1.021958 &   29.340847 &  1.188433\\
 &    1.259114 &  1.433044 & -155.301234 &  1.488433\\
 &   -2.105286 &  1.844130 &  201.332436 &  1.788433\\
 &    1.051255 &  2.255216 &  -78.899791 &  2.088433\\
      \hline
$^3d_1D$ &    0.392036 &  0.756905 &    0.004056 &  0.394786\\
Term 1  &    9.293162 &  1.399407 &    0.094926 &  0.825067\\
 &  -54.630929 &  2.041908 &   -0.620590 &  1.255347\\
 &   74.598555 &  2.684410 &    0.869670 &  1.685627\\
 &  -29.652824 &  3.326911 &   -0.348062 &  2.115907\\
      \hline
Term 2  &    0.018926 &  0.950962 &   34.649504 &  0.697745\\
 &   -0.178877 &  1.526973 & -323.198246 &  1.098823\\
 &    0.468087 &  2.102984 &  846.220442 &  1.499901\\
 &   -0.466509 &  2.678995 & -845.312868 &  1.900980\\
 &    0.158373 &  3.255006 &  287.641168 &  2.302058\\
\end{tabular}
\end{table}

\begin{table}[ht]
\caption{Results for the $\pi ^3$He scattering length.
The first three entries contain measured energy level shifts of
$\pi ^3$He atomic bound states together with corresponding
scattering lengths extracted by using Eq. (\protect{\ref{scl}}).
The averaged values includes an additional error of 10 \% to take
into account possible effects from isospin breaking, etc., cf.
text.}
\label{scltab}
\begin{center}
\begin{tabular}{lcc}
&$\epsilon_{1s}$ [eV] & $a_{\rm He}$ [$m_\pi^{-1}$] $\times 10^3$\\
\hline
R. Abela {\it et al.} \protect{\cite{phe1}} &
$44 \pm 5$ & $56 \pm 6$ \\
G. R. Mason {\it et al.} \protect{\cite{phe2}} &
$34 \pm 4$ & $43 \pm 5$ \\
I. Schwanner {\it et al.} \protect{\cite{phe3}} &
$32 \pm 3$ & $41 \pm 4$ \\
averaged value && $47 \pm 8 \pm 5$ \\
\hline
theoretical prediction && $67 \pm 15$
\end{tabular} \end{center}
\end{table}

\newpage

\end{document}